\documentclass[
    ,final            
  ]
  {aipproc}

\layoutstyle{6x9}

\begin{document}

\title{Modeling Emission from the First Explosions:  Pitfalls and Problems}


\author{Chris L. Fryer}{
  address={CCS-2, MSD409, Los Alamos National Laboratory, Los Alamos, NM 87545}
}

\author{Daniel J. Whalen}{
  address={Department of Physics, Carnegie Mellon University, Pittsburgh, PA 15123}
}

\author{Lucy Frey}{
  address={XTD-6, MSD409, Los Alamos National Laboratory, Los Alamos, NM 87545}
}

\begin{abstract}

Observations of the explosions of Population III (Pop III) stars 
have the potential to teach us much about the formation and evolution 
of these zero-metallicity objects.  To realize this potential, we must 
tie observed emission to an explosion model, which requires accurate
light curve and spectra calculations.  Here, we discuss many of the
pitfalls and problems involved in such models, presenting some
preliminary results from radiation-hydrodynamics simulations.

\end{abstract}

\maketitle


\section{Introduction}

JWST will advance our current picture of first-generation stars, which 
is now based on theory and indirect constraints from nucleosynthetic yields, 
by directly observing the explosions of these stars.  Such observations 
will vastly improve our understanding of the structures and masses of 
Pop III stars.  How much information these observations provide depends 
upon the accuracy of our simulations of the emission from primordial
supernovae.  In this paper, we discuss the problems and pitfalls that 
can befall such models.

Recent supernova light curve calculations have focused on thermonuclear 
explosions of white dwarfs: so-called Type Ia supernovae.  The usual 
approach in these models is to first simulate the explosion in a purely 
hydrodynamic calculation.  The trajectories from this explosion are then 
used in a radiation transport calculation.  The transport calculation 
assumes that an equilibrium between radiation and matter exists and 
allows this equilibrium to set the matter temperature based on the 
radiation flow due to the decay of radioactive nickel.  

This approach, separating the hydrodynamic and radiation transport
calculations, will not work for the first stars.  Indeed, we know 
that radiation-only calculations of light curves for all massive star 
supernovae are incorrect.  This is because the light curve is 
not just powered by radioactive decay.  A second heat source exists - 
that of shock heating.  Depending upon the initial radius of the 
star and the immediate surroundings of the explosion, shock heating 
can dominate the light curve out to late times~\cite{fryer07,fryer09}.  
Including the effects of shock heating in a supernova emission model 
requires a coupled radiation-hydrodynamics calculation.

As we shall show here, accurate estimates of supernova emission require 
modeling the departure of radiation and matter from equilibrium.  This 
departure is most pronounced when the radiation in the shock 
first becomes untrapped: at shock emergence or shock breakout.  The
shock breakout signal dominates the observable transient in first
stars, making accurate models of this nonequilibrium process
crucial for understanding detections of these outbursts.  In this 
paper, we discuss the physics of shock breakout and light curve
calculations, concluding with the current LANL approach for a new
generation of light curve models.

\section{Pitfalls in Modeling Shock Breakout}

\begin{figure}
  \includegraphics[height=.35\textheight]{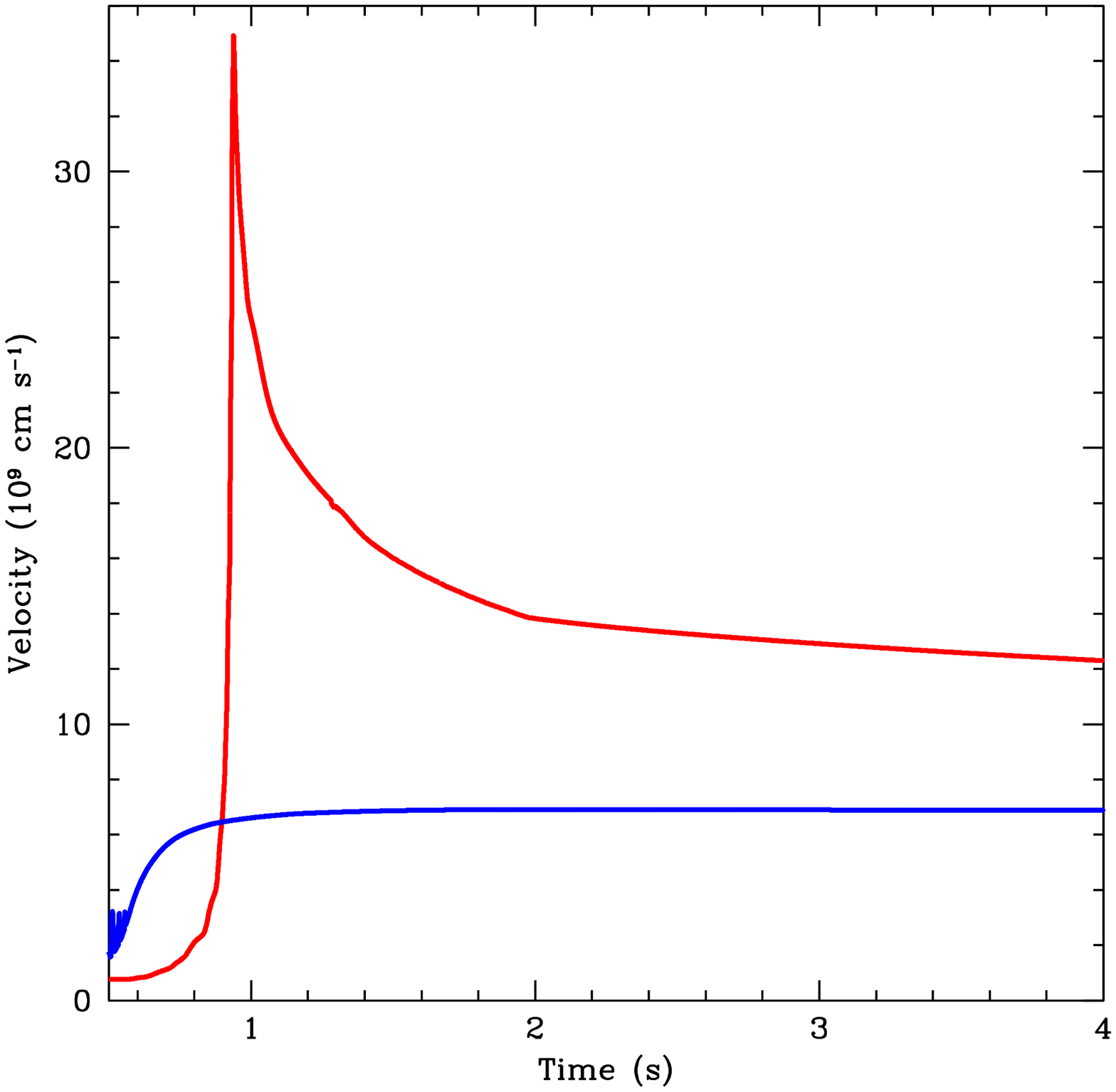}
  \includegraphics[height=.35\textheight]{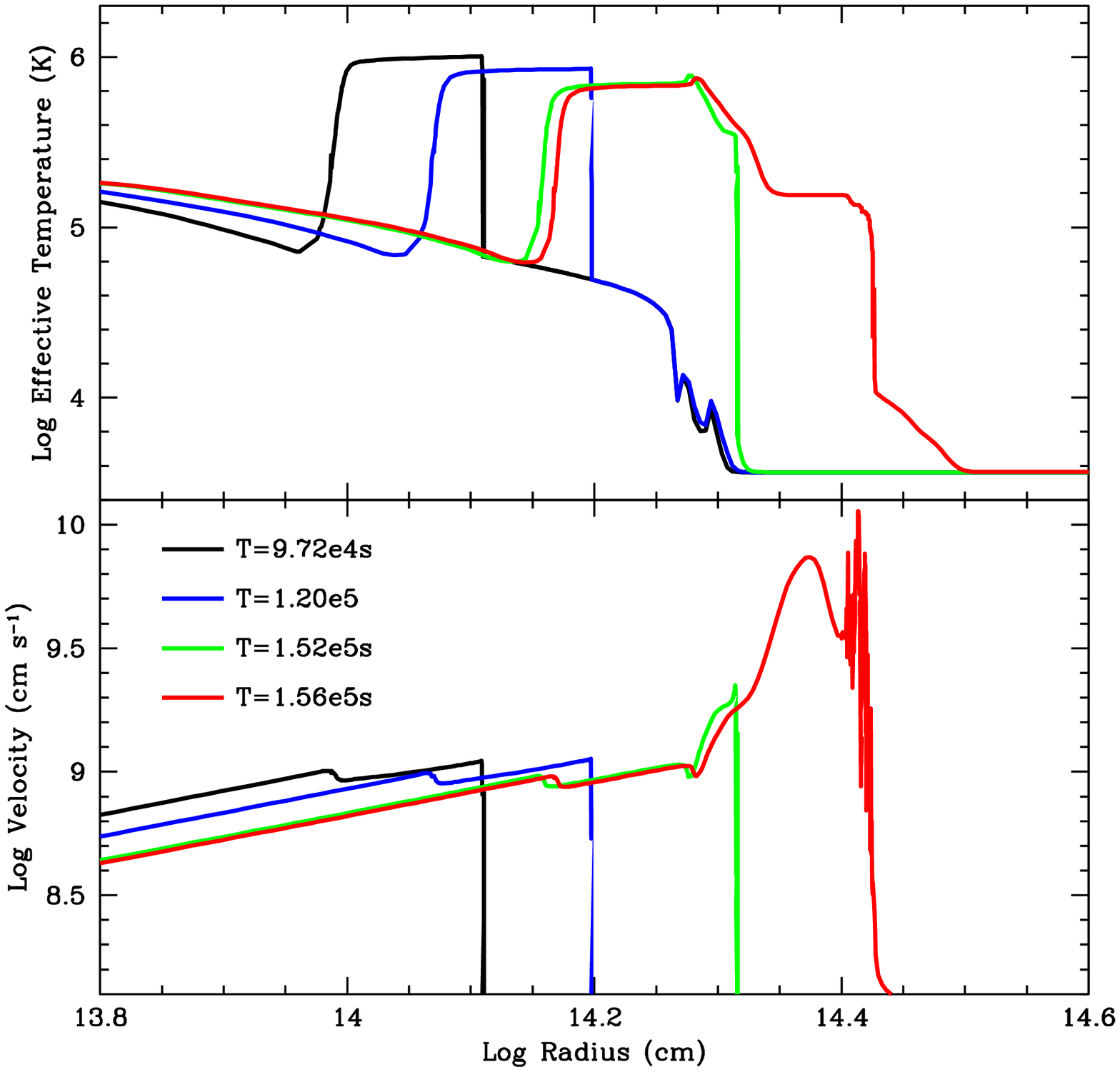}
  \caption{{\it Left:} Maximum shock velocity versus time in two
    models of a Wolf-Rayet star supernova: a purely
    hydrodynamic calculation with no radiative losses and a 
    simulation with radiative losses in a single-group,
    flux-limited diffusion model.  In the hydrodynamic-only
    calculation, the shock accelerates dramatically as it crosses
    the sharp density gradient at the edge of the star, reaching
    speeds above the speed of light in this non-relativistic
    model (meaning that the flow is relativistic with
    high Lorentz factors - shock breakout was once argued as a way to
    produce gamma-ray bursts~\cite{colgate74}).  It is clear from the 
    transport calculation that radiative
    losses diminish this acceleration dramatically.  Even at late
    times, well after shock breakout, the maximum velocities in these
    two models are very different.  {\it Right:} Effective radiation
    temperature defined by $(E_{rad}/\sigma)^{1/4}$ (top) and
    velocity (bottom) versus radius for 4 snapshots in time during
    shock breakout in a pair-instability supernova explosion.  In the
    initial two snapshots, the radiation is trapped in the flow of the
    shock.  By the third snapshot, the radiation has begun to
    lead the shock but is still coupled strongly with the
    material ahead of the shock and preaccelerates it.  In this manner,
    the radiation flow, not the hydrodynamic flow, determines the
    position and velocity of the shock.  The position of the shock
    changes faster than the shock velocity/Sedov solution would
    predict.  In the last snapshot, the radiation is becoming
    decoupled completely from the matter and the radiation front is no
    longer able to accelerate the matter.  This is the true escape of
    the radiation.  Note that radiation decoupling effects span a
    range of time with a variety of effects, and that one can not simply
    mimic these effects with a cooling term.}
\label{fig-fryer1}
\end{figure}

The engine for both core-collapse and pair-instability 
supernovae is launched in the stellar core.  While it is deep 
within the photosphere, the radiation is trapped in the 
flow, carried outward with the shock.  But as the shock 
breaks out of the star, the radiation begins to lead the 
shock and ultimately decouples from it.    

{\it Energy Loss:} In a purely hydrodynamic simulation, the explosion
is well described by the Sedov similarity solution~\cite{sedov59}: 
$v_{\rm shock} \propto t^{({\omega - 3})/(5-\omega)}$ where $\omega$ 
defines the density structure of the medium: $\rho \propto r^{-\omega}$.  
As the shock reaches the edge of the star, the density drops rapidly 
($\omega$ becomes large) and the shock can accelerate dramatically (as 
seen in the purely hydrodynamic calculation in the left panel of 
Fig.~\ref{fig-fryer1}).  But this calculation assumes that all of the 
internal energy in the shock is able to accelerate the material in 
front of it.  If some internal energy, and hence shock pressure, is 
released, the acceleration will be much lower.  This is exactly what 
happens when the radiation decouples from the matter.  In our sample 
Wolf-Rayet simulation (Fig.~\ref{fig-fryer1}, left), radiative losses 
sap the energy in the shock, and acceleration by pressure in the sharp 
density gradient is heavily reduced.  The maximum material velocity is 
much lower than what would be expected from the Sedov solution (or a 
purely hydrodynamic calculation).  Clearly, a pure hydrodynamic 
calculation does not produce the accurate flows needed for these light 
curve calculations.

{\it Radiative Acceleration:} If the only effect of radiation on the
hydrodynamical evolution were energy loss, it could be modeled with a
cooling term in the explosion calculation.  But radiative effects in 
shock breakout are much more complex than just energy loss.  As the 
radiation begins to decouple from the shock, it interacts with the 
matter in front of the shock strongly enough to accelerate it (right 
panel of Fig.~\ref{fig-fryer1}). The position of the shock will thus
be further out than predicted by a purely hydrodynamic model with a 
cooling term.  Radiation, not matter pressure gradients, governs the 
position and velocity of the outmoving shock.  The maximum velocity 
of this radiation-driven shock can also be faster than the Sedov 
solution would predict (counteracting the effect of radiative losses).  
This is visible in the early rise in the shock velocity in the sample
Wolf-Rayet simulation (Fig.~\ref{fig-fryer1}, left panel).

As the radiation further decouples, it begins to sap the shock strength 
(see our {\it Energy Loss} discussion), leading to lower velocities.  
These two radiative effects compete:  radiative acceleration causes the 
shock to move faster than expected from a purely hydrodynamic calculation 
and radiative losses cause the shock to move more slowly.  Radiation 
hydrodynamics is needed to capture both effects.

\section{Light Curve Models and Opacities}

\begin{figure}
  \includegraphics[height=.3\textheight]{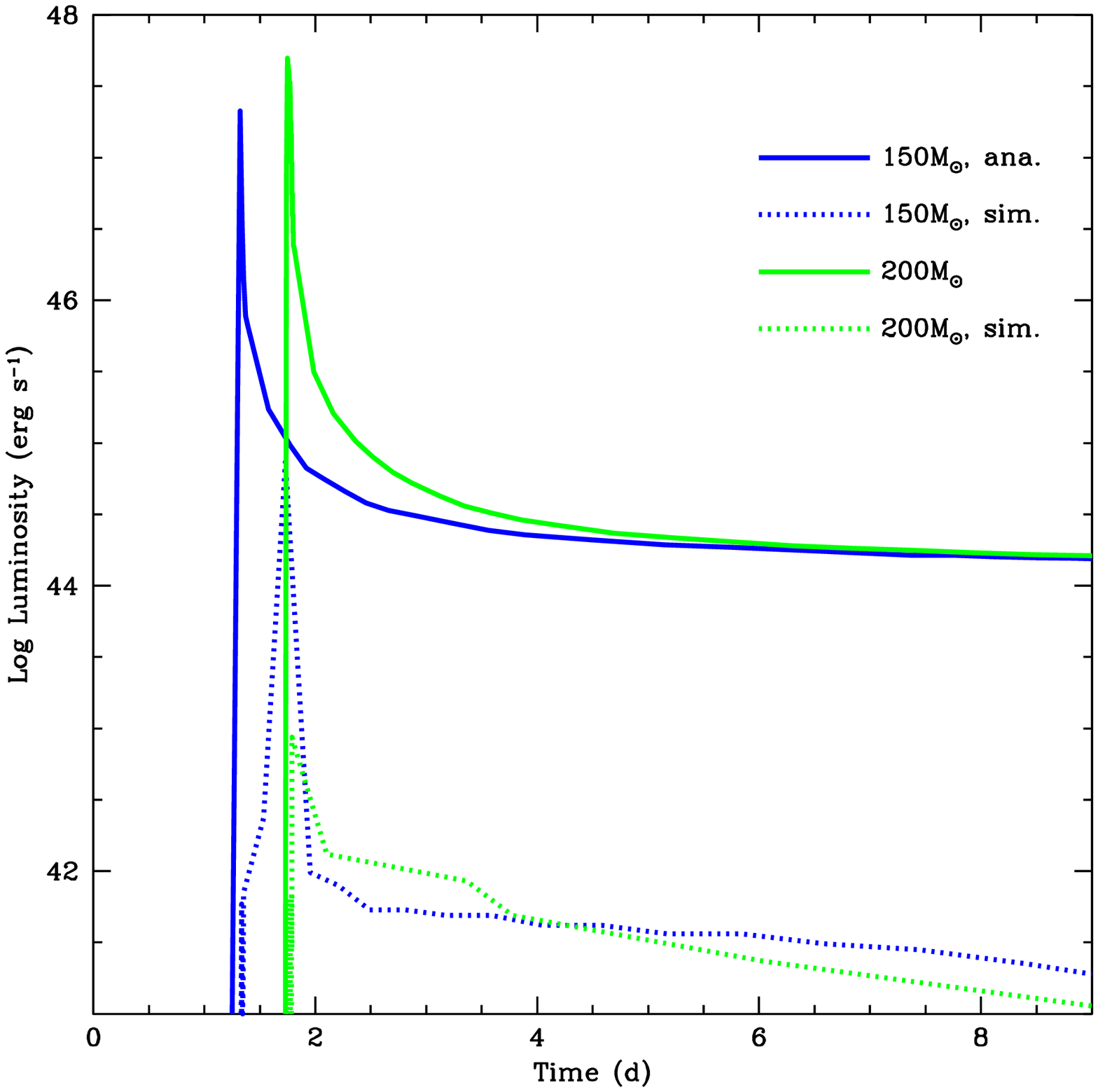}
  \includegraphics[height=.3\textheight]{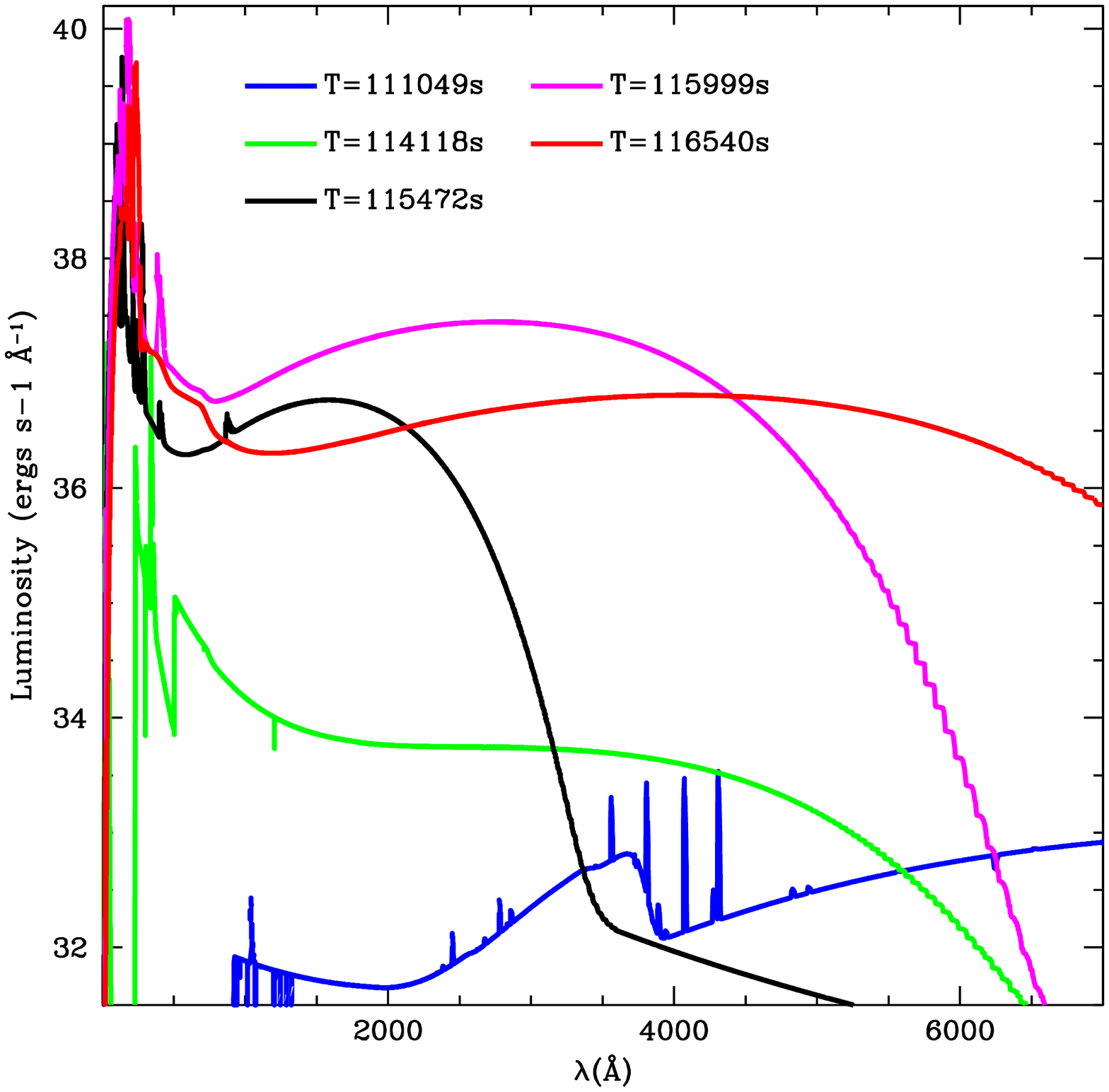}
  \caption{{\it Left:} Bolometric light curves (luminosity vs. time)
    of shock breakout for 2 different stars (dark line -
    150\,M$_\odot$ model, light line - 200\,M$_\odot$) as predicted by
    the standard analytic approach (solid) and by a full calculation.  For
    the standard analytic approach, we calculated the position of the
    photosphere and the temperature at this position, assuming an
    electron-dominated opacity ahead of the shock.  The bolometric
    luminosity was calculated assuming a blackbody source at this
    temperature and position.  In our simulated luminosity, we
    calculated the true emission using frequency dependent opacities.
    Note that the peak in calculated emission is over an order of magnitude
    less than that of the simple analytic estimate.  {\it Right:} Spectra
    (luminosity vs. wavelength) for a 150\,M$_\odot$ pair-instability
    supernova at 5 snapshots in time during shock breakout.  A broad
    set of emission and absorption lines are present during this
    process.  The initial burst of photons is at shorter
    wavelengths.  As the shock expands and cools, the emission at
    low wavelengths increases.
}

\label{fig-fryer2}
\end{figure}

A simple way to compute supernova light curves during shock breakout
is to calculate the location of the photosphere, determine the
temperature at the photosphere, and estimate the bolometeric
luminosity by assuming blackbody radiation: $L_{\rm SN} = 4 \pi
r_{\rm ph}^2 T_{\rm ph}^4$.  Depending upon the mass 
loss in the wind, the photosphere can be at the edge of the star or
deep within the star's wind profile.  Many theoretical estimates of
the bolometric luminosity use such a simple approximation.  Comparing
this analytic estimate to an actual simulation in which a full spectrum 
is calculated and then summed to get a bolometric luminosity reveals
that the analytic approach can overestimate the bolometric luminosity
by over an order of magnitude (Fig.~\ref{fig-fryer2}, left).

The crude estimate of the opacity in the analytic approach accounts
for a large fraction of this discrepancy.  The photosphere is not 
at a single location, but depends on wavelength and the wavelength 
dependent opacity.  Because the luminosity is proportional to $T^4$ 
and the temperature gradient is so steep at breakout, small errors 
in the shock position can produce large changes in the total 
predicted luminosity.  Detailed opacities as a function of wavelength 
are crucial for accurate breakout luminosity 
predictions.  Another assumption typically used in computing 
opacities is that the atomic level states are in equilibrium with 
the matter temperature. In shock breakout, this is not necessarily 
true and non-LTE effects may alter the opacities, and hence 
luminosities.  

\section{LANL Approach}

LANL has begun to leverage off of its expertise in radiation
hydrodynamics techniques as part of the Advanced Simulation and
Computing (ASC) program.  Under this program, LANL has developed 
the multi-dimensional radiation-hydrodynamics code RAGE (Radiation
Adaptive Grid Eulerian), which was designed to model multi-material
flows~\cite{gittings08}.  RAGE is an adaptive mesh refinement (AMR)
code with a second-order, direct-Eulerian Godunov hydrodynamics 
solver and multi-group flux-limited diffusion radiation transport 
with operator-split coupling. A discrete-ordinate scheme has been 
added to model non-thermal photon transport~\cite{fryer09} along 
with implicit Monte Carlo (IMC) transport with full angular 
information for thermal photons.

There is some confusion in the astrophysical literature between the
``diffusion'' schemes used in most stellar evolution codes and the 
flux-limited diffusion (FLD) used in transport.  Diffusion schemes 
in stellar models tend to be akin to conduction schemes, allowing 
energy flow from hotter to colder zones that is determined by the 
temperature gradient and the radiative diffusion coefficient.  Such 
equilibrium diffusion/conduction schemes are appropriate in optically 
thick media where the radiation is fully coupled to the matter.  FLD
schemes model the radiation flow separately from the matter (the 
radiation energy is not set by the matter temperature).  The 
simplifying assumption in these schemes is that the angular 
distribution of the radiation has a specific form.  Full, or 
higher-order, transport schemes make no assumptions about the angular 
distribution but instead calculate it. The FLD
transport in RAGE can model the decoupling of the radiation field from 
the shock in our breakout calculations.  

Some astrophysical implementations of higher-order transport have made
assumptions that the radiation and matter are fully coupled.  Although
these schemes include the angular distribution of the radiation field,
they are little better than conduction codes in modeling shock breakout.  
These are sometimes referred to as 1-temperature algorithms, while codes 
that evolve radiation flux and matter temperature separately are known 
as, somewhat confusingly, 2-temperature codes.  It is important to know
what scheme is used when analyzing results from theory calculations.  At 
LANL, all our transport algorithms (including our discrete ordinate and 
Monte Carlo methods) are full 2-temperature schemes.

At present, we assume that the level populations of the atoms in our 
shocks are in equilibrium with the matter.  If the radiation field is 
exciting these atoms, this thermodynamic equilbrium is certainly not 
valid, but the database of nonequilibrium opacities are not yet 
sufficiently complete for our calculations (LANL's atomic physicists 
are currently working on such a database).  In addition, the modeling 
of radiation flow during shock breakout is very sensitive to the 
operator-split coupling term and it is likely that our current scheme 
is introducing errors.

\begin{figure}
  \includegraphics[height=.35\textheight]{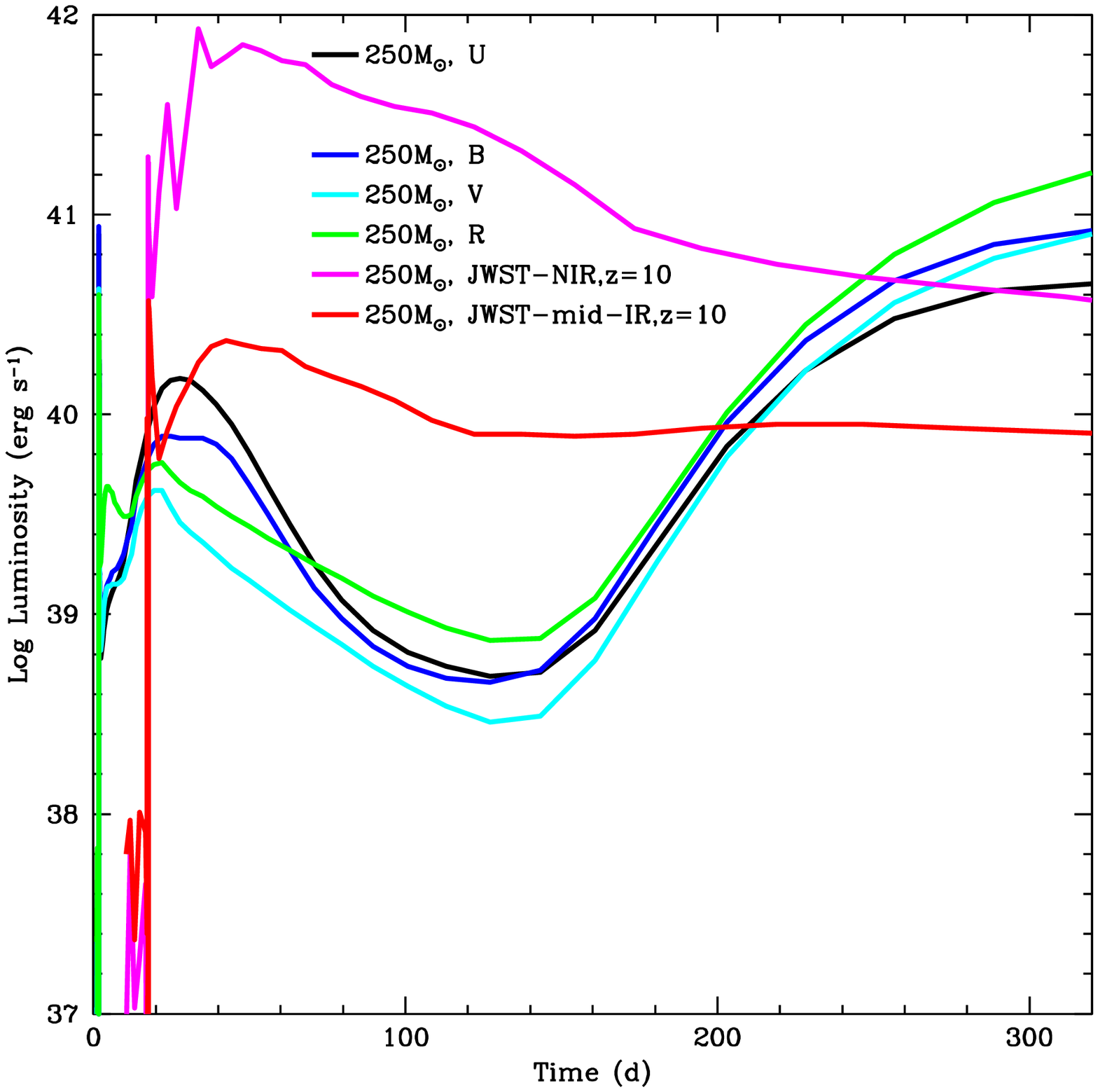}
  \includegraphics[height=.35\textheight]{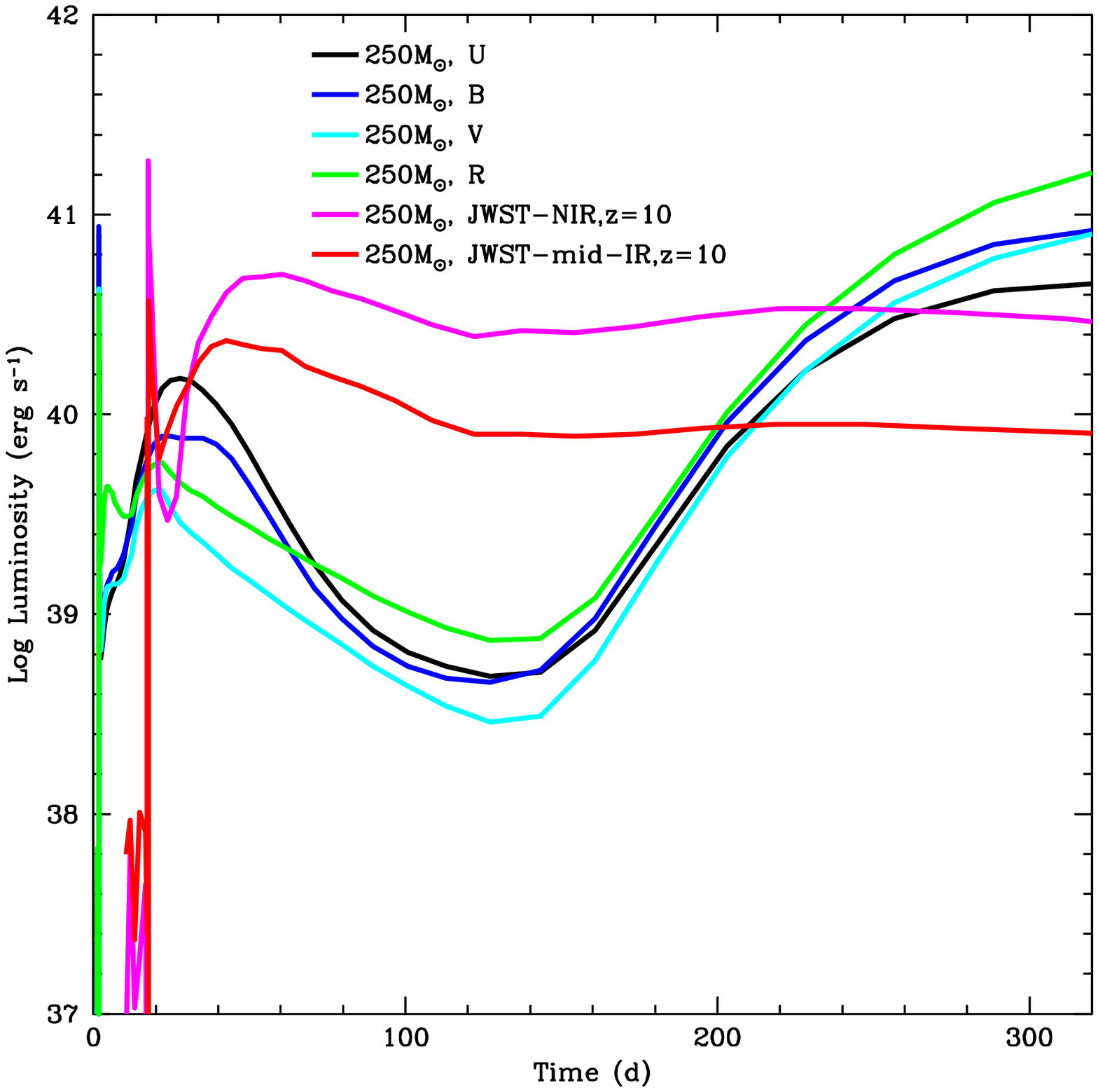}
  \caption{Light curves (luminosity versus time) for a number of
    spectral bands for a 250 M$_{\odot}$ pair-instability supernova 
    out to 1 year.  We have included two bands consistent with the 
    JWST bands assuming that the source is at a redshift of 10. Here, 
    we have included both the wavelength shift of the source emission 
    and time dilation.  Note that the shock breakout emission lasts a
    week in the JWST frame.  The differences between the plots on the 
    left and on the right is the inclusion of the Lyman break, which
    cuts off any emission shortward of 1216\AA.  Most high-redshift 
    supernova searches for these pair-instability supernovae will detect 
    the shock breakout.}
\label{fig-fryer3}
\end{figure}

With these caveats, we conclude with a brief discussion of our initial 
results.  Figure~\ref{fig-fryer2} (right) shows the wealth of spectral 
data in the shock breakout light curves for a 250\,M$_\odot$ 
pair-instability explosion.  Both emission lines and absorption lines 
(especially in the ultra-violet) are present. Many of these lines will 
appear in the visible spectrum at high redshift. We show the light
curves of this supernova in a variety of bands in Fig.~\ref{fig-fryer3}.  
At low energies, the peak in the light curve can occur a year after the
explosion.  Redshifted to first stars, this peak would occur 10 - 20 
years after the explosion.  Such a supernova would not be detected as a
transient.  The shock breakout itself will exhibit a broad peak lasting 
up to 200 days.

The prolonged pair-instability light curves, coupled with high-redshift 
time dilation effects, limit the potential of high-redshift transient 
surveys.  JWST may only detect the shock breakout emission.  Modeling 
this emission not only requires true 2-temperature radiation-hydrodynamic 
calculations, but probably nonequilibrium opacities as well. Until all of 
this is done accurately, any theoretical predicition must be taken with 
some grain of salt.

\begin{theacknowledgments}
This work was carried out in part under the auspices of the National
Nuclear Security Administration of the U.S. Department of Energy at
Los Alamos National Laboratory supported by Contract
No. DE-AC52-06NA25396. DJW was supported by the McWilliams Fellowship 
at the Bruce and Astrid McWilliams Center for Cosmology at Carnegie 
Mellon University.

\end{theacknowledgments}



\bibliographystyle{aipproc}   


\begin{thebibliography}{4}

\bibitem{colgate74}
Colgate, S.A., \emph{ApJ}, \textbf{187}, 333--336 (1974)

\bibitem{sedov59}
L.~I. Sedov, Similarity and Dimensional Methods in Mechanics, 
Academic Press, New York, 1959.

\bibitem{fryer07}
Fryer, C.L., Hungerford, A.L., \& Young, P.A., \emph{ApJ Lett}, \textbf{662}, 
55-58, (2007)

\bibitem{fryer09}
Fryer, C.L., et al., \emph{ApJ}, \textbf{707}, 193--207 (2009)

\bibitem{gittings08}
Gittings, et al., \emph{Comp. Science and Discovery}, \textbf{1}, 015005 (2008)

\end{thebibliography}



\end{document}